\documentclass[preprint,showpacs,preprintnumbers,amsmath,amssymb]{revtex4}
\usepackage{graphicx}
\usepackage{dcolumn}
\usepackage{bm}

\newcommand{\be}{\begin{equation}}
\newcommand{\en}{\end{equation}}
\newcommand{\bea}{\begin{eqnarray}}
\newcommand{\ena}{\end{eqnarray}}

\begin{document}


\title{Warped DGP model in warm intermediate inflation with a
general dissipative coefficient in light of BICEP2 and Planck
results }
\author{Ram\'on Herrera}

\email{ramon.herrera@ucv.cl} \affiliation{ Instituto de
F\'{\i}sica, Pontificia Universidad Cat\'{o}lica de Valpara\'{\i}so, Avenida Brasil 2950,
Casilla 4059, Valpara\'{\i}so, Chile.}

\author{Marco Olivares}
\email{marco.olivaresr@mail.udp.cl} \affiliation{ Facultad de
Ingenier\'{\i}a,
 Universidad Diego Portales,  Avenida Ej\' ercito Libertador 441, Casilla
298-V, Santiago, Chile.}

\author{Nelson Videla}
\email{nelson.videla@ucv.cl} \affiliation{ Instituto de
F\'{\i}sica, Pontificia Universidad Cat\'{o}lica de
Valpara\'{\i}so, Avenida Brasil 2950, Casilla 4059, Valpara\'{\i}so, Chile.}

\date{\today}

\begin{abstract}
 A warm inflationary universe scenario on a warped Dvali-Gabadadze-Porrati
 brane during intermediate inflation is studied.  We consider  a
 general form for the dissipative coefficient
$\Gamma(T,\phi)\propto T^{m}/\phi^{m-1}$, and also study this
model in the weak and strong dissipative regimes. We analyze the
evolution of  the universe in the slow-roll approximation, and
find the exact solutions to the  equations of motion. In both
regimes, we utilize recent data from the BICEP2 experiment and also from the Planck satellite to
constrain the parameters in our model in accordance with the theory of
cosmological perturbations.
\end{abstract}

\maketitle

\section{Introduction}

It is well known that the scenario of warm inflation is different
than that of traditional cold inflation, notably in that warm inflation avoids
a reheating period \cite{warm}. During the evolution of warm
inflation dissipative effects are important, and radiation
effects take place at the same time as the  expansion of the
universe. The dissipating effect occurs due to a friction term
which accounts for the processes of the scalar field dissipating into
a thermal bath. In further relation to these dissipative effects,
the dissipative coefficient $\Gamma$ is a fundamental quantity. This parameter
$\Gamma$ was studied in a supersymmetric model \cite{26} for low-temperature
scenario. For a scalar field
with multiplets of heavy and light fields that give different
expressions for the dissipation coefficient, see
Refs.\cite{26,28,2802,Zhang:2009ge,BasteroGil:2011xd,BasteroGil:2012cm}.
A general form for the dissipative coefficient
$\Gamma(T,\phi)$, can be written as
\cite{Zhang:2009ge,BasteroGil:2011xd}

\begin{equation}
\Gamma(T,\phi)=C_{\phi}\,\frac{T^{m}}{\phi^{m-1}}, \label{G}%
\end{equation}
where  the constant  $C_\phi$ is associated  with  the dissipative
microscopic dynamics and the constant $m$ is an integer. Various
values of $m$ have been considered in the literature, see Refs.
\cite{Zhang:2009ge,BasteroGil:2011xd}. Specifically, for the value
of $m=3$, i.e., $\Gamma \propto T^3\phi^{-2}$, the parameter
$C_\phi$ corresponds to  $C_{\phi}=0.02\,h^{2}\,\mathcal{N}_Y$,
where a generic supersymmetric model with chirial superfields
$\Phi$, $X$ and $Y_i$, $i=1,...\mathcal{N}_Y$ is considered
\cite{BasteroGil:2012cm,new}. For the special value $m=1$,
$\Gamma$ is associated with the high temperature supersymmetry
(SUSY) case. For the special case $m=0$, the dissipation
coefficient represents an exponentially decaying propagator in the
high temperature SUSY model. For the value $m=-1$, i.e.,
$\Gamma\propto\phi^2/T$, agrees with the non-SUSY
case\cite{28,PRD}. Additionally, thermal fluctuations during the
inflationary scenario may play a fundamental  role in producing
the initial fluctuations essential for Large-Scale Structure (LSS)
formation\cite{62526,1126}.  During the warm inflationary scenario
the density perturbations arise from the thermal fluctuations of
the scalar field  and dominate over the quantum origin of the
initial density perturbations. In this form, an essential
condition for warm inflation scenario is the existence of a
radiation component with temperature $T>H$, during the expansion
of the universe, since the thermal and quantum fluctuations are
proportional to $T$ and $H$, respectively\cite{warm,62526,1126}.
Also relevant, as the universe heats up and becomes radiation
dominated, then warm inflation ends. Here, the universe stops
inflating and smoothly enters   in a radiation Big-Bang
phase\cite{warm}. For a comprehensive review of warm inflation,
see Ref. \cite{Berera:2008ar}.

On the other hand, from high dimensional gravity theory,
Dvali-Gabadadze-Porrati (DGP) considered a braneworld
model\cite{DGP} where the universe is a 4-dimensional brane
embedded in a  5-dimensional Minkowski space-time. In this perspective, the
induced gravity brane-world in the DGP model was put forward as an
alternative to the Randall-Sundrum (RS) one-brane model \cite{RS}.
The gravitational behaviors in the DGP model are divided between
the 5-dimensional curvature scalar in the bulk and the
4-dimensional curvature scalar on the brane. According to the
embedding of the brane in the bulk in the DGP brane, there are two
branches of background solutions, i.e, there are two forms to
embed space-time (the 4-dimensional brane into the
5-dimensional space). The inflationary universe model in the
context  of warped DGP has been analyzed in Refs.\cite{Bou,Pap,
Rong, Rong2,Herrera:2013rca}. In particular,  the
warm inflation model in a DGP scenario was analyzed  in Ref.\cite{del
Campo:2007zj}, where the authors studied  a standard scalar field
coupled to radiation for an effective exponential potential. Also,
the cosmological dynamics of a quintom field on the warped DGP
brane was analyzed in Ref.\cite{Nozari:2014kpa}, and observational
constraints on the normal branch of a warped DGP cosmology were
found in Ref.\cite{Azizi:2011ys}.  For a review of the DGP model,
see Ref.\cite{Lue}.

On the other hand, in the context of the exact solutions,
can be obtained for instance in the intermediate inflation model \cite{Barrow1}, where
the scale factor is slower than de-Sitter expansion, but quicker
than power law (hence the name ``intermediate"). During this
scenario, the scale factor grows as
\begin{equation}
a(t)=\exp[\,A\,t^{f}],  \label{at}
\end{equation}
where $A$ and $f$ are two constants; and in which  $A>0$ and the
constant $f$ varies between $0<f<1$ \cite{Barrow1}. It is well
known  that the exact solutions in inflationary scenarios can be
obtained from an exponential potential or  "power-law" inflation,
in which   $a(t)\sim t^{p}$, where $p>1$\cite{power}.  Similarly,
an exact solution can be found from a constant potential, "de
Sitter" inflation\cite{R1}. In the context of the intermediate
model, the scale factor $a(t)$  was earlier elaborated as an exact
result of the equation of motion, but the dynamics of the model  may be
best described from slow-roll approximation. In the slow-roll
approximation, it is feasible to obtain a spectral index $n_s\sim
1$ (and in particular for the specific  value of $f=2/3$, the
spectral index results  $n_s=1$, or the Harrizon-Zel'dovich
spectrum). Likewise, an important observational magnitude found in
the intermediate model is the tensor to scalar ratio $r$,
which becomes $r\neq 0$\cite{Barrow3}.

Thus the aim  of the  article is to study an intermediate scale
factor during  warm  inflation scenario in the framework of a warped DGP
model, and how  a generalized form of
dissipative coefficient $\Gamma(T,\phi)\propto T^{m}/\phi^{m-1}$ influences our model.
We will consider a warm intermediate inflation on a warped DGP
model for two regimes; the weak and the strong
dissipative scenarios, respectively. Also, we will investigate the
cosmological perturbations, which are expressed in terms of
different parameters appearing in our model. These parameters are
constrained by the  BICEP2 experiment data\cite{B2} and the Planck satellite\cite{Planck}. The  BICEP2
results
 imply a large amplitude of primordial gravitational
waves and hence  has important  theoretical significance  on
 inflationary universe models. The observational data on the
tensor-to-scalar ratio $r$, has been obtained at more than
5$\sigma$ confidence level (CL) with a rigorous constraint, where
$r = 0 . 20^{ +0 . 07}_{ - 0 . 05}$ at 68 $\%$ CL, also  $r = 0 .
16 ^{+0 . 06}_{ - 0 . 05}$ with foreground subtracted. However,
the ratio $r$  has become less clear when grave criticisms to the
BICEP2 analysis appeared in the literature. Recently, the Planck
collaboration has issued the data relating the polarized dust
emission \cite{Adam:2014bub}. Here, an analysis of the polarized
thermal emissions from diffuse Galactic dust in the range from 353
GHz to 150 GHz, suggests that BICEP2 gravitational wave result
could be due to the dust contamination, and a detailed study of
Planck and BICEP2 data would be required for a definitive answer.

The outline of the paper is as follows. The next section presents a
short review of the Friedmann equation on the warped DGP inflation
model. In Section \ref{secti} we present the warm inflationary
phase on the warped DGP model, study the weak and strong dissipative regimes, and discuss the inflationary epoch and
the cosmological pertubations in both regimes.
 Finally, Sec.\ref{conclu}
summarizes our findings. We chose units so that $c=\hbar=1$.


\section{ The Friedmann equation on the warped DGP brane}

The Friedmann equation on the warped DGP model can be written from
the Friedmann-Robertson-Walker (FRW) metric as
 \be
 \label{H1}
 H^2={1\over 3\mu^2}\Bigl[\,\rho+\rho_0\bigl(1 +
\epsilon{\cal A}(\rho, a)\bigr)\,\Bigr],
 \en
 where   $H=\dot{a}/a$ corresponds to the Hubble parameter and
 $\rho$ is the total energy density. Here, the dots mean derivatives with respect to
time.  The constant $\mu$  denotes the strength of the induced
gravity term on the
 brane (the special case when $\mu=0$ yields the RS model\cite{RS}).  The parameter $\epsilon$  corresponds to  $+1$ or $-1$ and
 are  the two branches of the  warped DGP brane. For the value $\epsilon=-1$,  we will consider the brane tension
 as positive, and for the value $\epsilon=+1$
 negative. Here, the function
   ${\cal A}(\rho, a)$ is
defined as

\be
 {\cal A}=\left[{\cal A}_0^2+{2\eta\over
\rho_0}\left(\rho-\mu^2 {{\cal E}_0\over a^4}
\right)\right]^{1\over 2},
 \en
where the constants ${\cal A}_0$, $\rho_0$ and $\eta$ are given
by
 $
 {\cal A}_0=\sqrt{1-2\eta{\mu^2\Lambda\over \rho_0}},\,\;\;\rho_0=m_\lambda^4+6{m_5^6\over
 \mu^2},\;\;
\eta={6m_5^6\over \rho_0\mu^2} ~~~(0<\eta\leq 1),
 $
 and the constant  $\Lambda$, becomes
 $
\Lambda={1\over 2} ({}^{(5)}\Lambda+{1\over
6}\kappa_5^4\lambda^2).
 $
 Here, $\kappa_5$ is the 5-dimensional Newton constant, ${}^{(5)}\Lambda$ corresponds to the 5-dimensional cosmological constant
 in the bulk, ${\cal E}_0$ is a constant related to Weyl radiation, and  the brane tension is specified  by $\lambda$.
 Here,  there are three mass scales, $\mu$,
$m_\lambda=\lambda^{1/4}$ and $m_5=\kappa_5^{-2/3}$. Since we are only concerned with inflationary dynamics in
our model, we will ignore  the dark radiation term. Also,
 we shall restrict ourselves to
 the RS critical
case, where $\Lambda=0$. In this form,  Eq.(\ref{H1}) yields
 \be
 \label{newfried}
 H^2=\frac{1}{3\mu^2}\left[\rho+\rho_0+\epsilon \rho_0 \left(1
 +\frac{2\eta\rho}{\rho_0}\right)^{1/2}\right].
 \en

 Note that in the ultra high energy limit in which
 $ \rho \gg \rho_0 \gg m_{\lambda}^4$,  the Friedmann equation given by  Eq.(\ref{newfried})
 becomes, $H^2\propto(\rho+\epsilon\sqrt{2\rho\,\rho_0})$
 and corresponds to 4-dimensional gravity on the brane.  Also, in the
 intermediate energy region in which $\rho \ll \rho_0 $ but $\rho \gg
 m_{\lambda}^4$, for the branch with $\epsilon =-1$, Eq.(\ref{H1}) becomes $H^2  \propto\left( \rho
+\frac{\rho^2}{2m_{\lambda}^4} -\frac{\mu^2
m_{\lambda}^4}{6m_5^6}\rho -\frac{\mu^2}{4m_5^6}\rho^2\right),$
and  at the low energy limit, where $\rho \ll m_{\lambda}^4 \ll
  \rho_0$ then  Eq.(\ref{H1}) becomes
  $H^2\propto[\rho+\mathcal{O}(\rho/\rho_0)^2]$,
 where  $\mu_p$ is the effective 4-dimensional Planck mass and  $\mu^2_p= \mu^2/(1-\eta)$.

\section{Warm Inflation: Basic equations.\label{secti}}

In the following, we consider the universe filled with a
self-interacting scalar field of energy density $\rho_{\phi}$
together with  a radiation field
given by $\rho_{\gamma}$.  In this form, the total energy density
$\rho$ can be written as $\rho=\rho_{\phi}+\rho_{\gamma}$. In the following, we
will assume that the energy density associated with the scalar
field is $\rho_{\phi}=\dot{\phi}^{2}/2+V(\phi)$ and the pressure
is $P_{\phi}=\dot{\phi}^{2}/2-V(\phi)$, where the quantity $V(\phi)$ represents
 the effective  potential.

Also, we will assume that the total energy density $\rho$  is confined
to the brane in the bulk satisfying the continuity equation given
by $\dot{\rho}+3H(\rho+P)=0$. In this form, the dynamical
equations for $\rho_{\phi}$ and $\rho_{\gamma}$, are described
by\cite{warm}
\begin{equation}
\dot{\rho_{\phi}}+3\,H\,(\rho_{\phi}+P_{\phi})=-\Gamma\;\;\dot{\phi}^{2},
\label{key_01}%
\end{equation}
and
\begin{equation}
\dot{\rho}_{\gamma}+4H\rho_{\gamma}=\Gamma\dot{\phi}^{2}, \label{key_02}%
\end{equation}
where the  dissipation coefficient $\Gamma>0$( see Eq.(\ref{G})),
and produces decay of the scalar field into radiation. Recall, that the
parameter $\Gamma$ can be assumed  to be a function of the
temperature of the thermal bath $\Gamma(T)$, or a function of the
scalar field $\Gamma(\phi)$, or a function of $\Gamma(T,\phi)$ or
simply a constant\cite{warm}.

In the context of warm inflation, the energy density related to
the scalar field predominates  over the energy density of the
radiation field, i.e.,
$\rho_\phi\gg\rho_\gamma$\cite{warm,62526,6252602,6252603,6252604}
 and  then $\rho\sim \rho_\phi$. In this approximation, Eq.(\ref{newfried}) can be written as
$$
H^{2}\approx \frac{1}{3\mu^2}\left[\rho_{\phi}+\rho_0+\epsilon \rho_0 \left(1
 +\frac{2\eta\rho_{\phi}}{\rho_0}\right)^{1/2}\right],
$$
or equivalently as
\begin{equation}
H^{2}\approx\frac{1}{3\mu^2}\left[\frac{\dot{\phi}}{2}+V(\phi)+\rho_0+\epsilon \rho_0 \left(1
 +\frac{\eta [\dot{\phi}+2 V(\phi)]}{\rho_0}\right)^{1/2}\right]. \label{inf2}%
\end{equation}

 From Eqs.(\ref{key_01}) and (\ref{inf2}),  we get

\begin{equation}
 \dot{\phi}^2= 2\mu^2{(-\dot{H})\over (1+R)}\left[1-
\epsilon\left(\alpha+\beta H^2\right)^{-1/2}
\right],\label{a4}
\end{equation}
where $R=\frac{\Gamma}{3H}$ denotes the rate between $\Gamma$ and
the Hubble parameter. Note that for the case of the weak or strong
dissipation regime, we make $R<1$ or $R>1$, respectively.
Following Ref.\cite{Herrera:2013rca}, the constants $\alpha$ and
$\beta$ are defined by
$$
\alpha= 1+\frac{{\cal{A}}_0 ^{2}}{\eta^2}-\frac{2}{\eta}
,\;\;\;\mbox{and} \,\,\; \beta=\frac{6\mu^2}{\eta\rho_0}.
$$

On the other hand,  during the inflationary scenario, we assume
that radiation production is quasi-stable, i.e., $\dot{\rho
}_{\gamma}\ll4H\rho_{\gamma}$ and $\dot{\rho}_{\gamma}\ll\Gamma\dot{\phi}^{2}%
$, see Refs.\cite{warm,62526,6252602,6252603,6252604}. In this
form, by using Eqs.(\ref{key_02}) and (\ref{a4}), the density $\rho_{\gamma}$,
becomes
\begin{equation}
\rho_{\gamma}=\frac{\Gamma\dot{\phi}^{2}}{4H}=\frac{\mu^2 \,\Gamma(-\dot{H})}{2 H(1+R)}
\left[1-\epsilon\left(\alpha+\beta H^2\right)^{-1/2}\right]. \label{rh}%
\end{equation}

Also, the energy density of the radiation field
could be written as $\rho_{\gamma}=C_{\gamma}\,T^{4}$, where the constant  $C_{\gamma}%
=\pi^{2}\,g_{\ast}/30$. Here,   $g_{\ast}$ represents the number
of relativistic degrees of freedom.  Combining Eq.(\ref{rh}) with
$\rho_{\gamma}\propto\,T^{4}$, we get
\begin{equation}
T=\left[\frac{\mu^2 \,\Gamma\,(-\dot{H})}{2\,C_{\gamma}H\,(1+R)}\right]
^{1/4}\left[1-\epsilon\left(\alpha+\beta H^2\right)^{-1/2}\right]^{1/4}. \label{rh-1}%
\end{equation}
In particular for the weak dissipative regime in which $R<1$, from
Eqs. (\ref{G}) and (\ref{rh-1}), the temperature of
the thermal bath $T$ becomes
\begin{equation}
T=\left(\left[\frac{C_\phi\,\mu^2\,\phi^{1-m}
\,(-\dot{H})}{2\,C_{\gamma}H}\right]
\left[1-\epsilon\left(\alpha+\beta
H^2\right)^{-1/2}\right]\right)^{\frac{1}{4-m}},\label{TD}
\end{equation}
and for the strong dissipative regime ($R>1$), the temperature
is given by
\begin{equation}
T=\left[\frac{3\mu^2 \,(-\dot{H})}{2\,C_{\gamma}\,H}\right]
^{1/4}\left[1-\epsilon\left(\alpha+\beta H^2\right)^{-1/2}\right]^{1/4}.\label{TF}
\end{equation}

 On the other hand,  the scalar potential $V$, can be found
 combining Eqs.(\ref{inf2}) and (\ref{a4})
$$
V=\frac{\eta\rho_0 }{2}\left(\alpha+\beta H^2\right) \left[1-
\epsilon\left(\alpha+\beta H^2\right)^{-1/2}
\right]^{2}-\frac{{\cal{A}}_0 ^{2}\rho_0}{2\eta}
$$

\begin{equation}
+\frac{\mu^2\,\dot{H}}{(1+R)}\,\left(1+\frac{3}{2}\,R\right)\left[1-
\epsilon\left(\alpha+\beta H^2\right)^{-1/2} \right] ,\label{a5}
\end{equation}
and  this effective potential, could also be obtained explicitly
in terms of the scalar field, i.e., $V=V(\phi)$.

Also, the dissipation coefficient, by using Eqs.(\ref{G}) and
(\ref{rh-1}),
 can be rewritten as
\begin{equation}
\Gamma^{{\frac{4-m }{4}}}\,(1+R)^{m \over 4}=C_{\phi}\left[
\frac{\mu^2}{2\, C_{\gamma}}\right] ^{m \over 4}\,\phi^{1-m} \left[
\frac{-\dot{H}}{H}\right]^{m \over 4}
\left[1-\epsilon\left(\alpha+\beta H^2\right)^{-1/2}\right]^{{m \over 4}}. \label{G1}%
\end{equation}
Note that Eq.(\ref{G1}) fixes the dissipation coefficient in
the weak (or strong) dissipative regime as a function of the
scalar field $\phi$ (or the cosmological time $t$).

In the following, we will consider our model for a dissipative
coefficient  $\Gamma=\Gamma(T,\phi)$ given by Eq.(\ref{G}), and we
will restrict ourselves to the weak and  strong dissipation
regimes.


\subsection{ The weak dissipative regime.\label{subsection1}}

Assuming  that our model evolves in agreement with the weak
dissipative regime, where $\Gamma<3H$, and using Eqs.(\ref{at})
and (\ref{a4}), we obtain

\begin{equation}
\phi(t)-\phi_0=\frac{F[t]}{K},\label{a7}
\end{equation}
where $\phi(t=0)=\phi_0$ is an integration constant, that can be
assumed $\phi_0=0$. The constant $K$ is defined by $ K\equiv a_f
\left(\frac{1-f}{2 \mu^2 Af} \right)^{1/2} (\beta  A^2
f^2)^{-a_f/2} $ and the function $F[t]$, corresponds to the
incomplete Lauricella function\cite{libro}, defined as
$$
F[t]\equiv\left(\alpha+\frac{\beta}{t^{2(1-f)}
}\right)^{\frac{-a_f}{2}}
F^{(3)}_{D}[a_f;1+\frac{a_f}{2},1+\frac{a_f}{2},\frac{-1}{2}
,a_f+1,\sqrt{\alpha},-\sqrt{\alpha},
\epsilon\left(\alpha+\frac{\beta}{t^{2(1-f)}
}\right)^{\frac{-1}{2}} ],
$$
where the constant $a_f$ is given by $ a_f=\frac{f}{2(1-f)}. $
 The Hubble parameter as a function of
the inflaton field, $\phi$, from Eq.(\ref{a7}), becomes $
H(\phi)=\frac{Af}{(F^{-1}[K\phi])^{1-f}}, $ where $F^{-1}$
represents the inverse function of the incomplete Lauricella
function\cite{libro}.

From Eq.(\ref{a5}), the scalar potential as function of the scalar
field becomes
\begin{equation}
V(\phi)\simeq \frac{\eta\rho_0 }{2}\left(\alpha+ \frac{\beta
A^2f^2}{(F^{-1}[K\phi])^{2(1-f)}}\right) \left[1-
\epsilon\left(\alpha+
 \frac{\beta A^2f^2}{(F^{-1}[K\phi])^{2(1-f)}}\right)^{-1/2}
\right]^{2} -\frac{{\cal{A}}_0 ^{2}\rho_0}{2\eta}.\label{a9}
\end{equation}
Here, we considered that  only the first term of Eq.(\ref{a5})
predominates at large values of scalar field $\phi$. Also, we
observed that we would have found the same scalar potential given
by Eq.(\ref{a9}) using the slow-roll approximations i.e.,
$\dot{\phi}^2\ll V(\phi)$ and $\ddot{\phi}\ll 3H\dot{\phi}$.

That said, introducing the dimensionless slow-roll
parameter $\varepsilon$, defined as  $
\varepsilon=-\frac{\dot{H}}{H^2}=\frac{1-f}{Af(F^{-1}[K\phi])^{f}},
$ and  with the  requirement  for inflation to take place,
$\varepsilon<$1 (or equivalently $\ddot{a}>0$), we get
$\phi>\frac{1}{K}F\left[\left(\frac{1-f}{Af}\right)^{1/f}\right]$.
Also, assuming that the inflationary scenario starts  at the
earliest possible scenario, in which $\varepsilon=1$ (see
Ref.\cite{Barrow3}), the field $\phi_1$ can be written as
$\phi_{1}=\frac{1}{K}F\left[\left(\frac{1-f}{Af}\right)^{1/f}\right]$.

Introducing the number of e-folds $N$ among two  values of
cosmological times $t_1$ and $t_2$, or equivalently among two
different values $\phi_1$ and $\phi_2$,  we get
\begin{equation}
N=\int_{t_1}^{t_{2}}\,H\,dt=A\,\left[(F^{-1}[K\phi_{2}])^{f}-(F^{-1}[K\phi_{1}])^{f}\right].\label{N1}
\end{equation}
Here we have considered  Eqs.(\ref{at}) and (\ref{a7}).

In the following, we will study  the scalar and tensor
perturbations in the weak dissipative regime for the warm warped
DGP brane. Following  Refs.\cite{warm,Liddle}, the density
perturbation is given by
${\cal{P}_{\cal{R}}}^{1/2}=\frac{H}{\dot{\phi}}\,\delta\phi$.
Here, we consider the gauge invariant quantity
$\zeta=H+\delta\rho/\dot{\rho}$, which is defined on slices of
uniform density and contracts to the curvature perturbation. A
characteristic  of this gauge invariant is that it  is closely
constant on super-horizon scales and does not depend on
gravitational dynamics \cite{P1} (see also, Ref. \cite{P2}). In
this case, the  spectrum associated with the curvature spectrum
could be written as ${\cal{P}_{\cal{R}}}^{1/2}\simeq
\sqrt{\langle\zeta^2\rangle}\simeq
\frac{H}{\dot{\phi}}\,\delta\phi$, which persists  unchanged in
the warped DGP model \cite{P3}.

To continue in the scenario warm inflation, a
thermalized radiation component exists and the fluctuations
$\delta \phi$ are predominantly thermal instead of quantum.
For the weak dissipative regime, the amplitude of the scalar field fluctuation
is given by $\delta\phi^2\simeq H\,T$ \cite{62526}. In this form, by using
Eqs.(\ref{a4}) and (\ref{rh-1}), the power spectrum
${\cal{P}_{\cal{R}}}$, results in
\begin{equation}
{\mathcal{P}_{\mathcal{R}}}={\sqrt{3\pi}\over 4\mu^2}\, \left(
\frac{\mu^2C_{\phi}}{2 C_{\gamma}}\right)
^{{\frac{1}{4-m}}}\phi^{{\frac{1-m}{4-m}}}H^{{\frac
{11-3m}{4-m}}}(-\dot{H})^{-{\frac{(3-m)}{4-m}}}\left[1-\epsilon\left(\alpha+\beta
H^2\right)^{-1/2}
\right]^{-{\frac{(3-m)}{4-m}}}. \label{pd1}%
\end{equation}

Also,  the power spectrum as function of the field $\phi$, from
  Eqs.(\ref{a7}) and (\ref{pd1}) can be written as

\begin{equation}
{\cal{P}_{\cal{R}}}(\phi)\simeq k_{1}\,\,\phi^{\,\frac{1-m}{4-m}}
\left(F^{-1}[K\phi]\right)^{\frac{2f(4-m)+m-5}{4-m}}\left[1-
\epsilon\left(\alpha+ \frac{\beta
A^2f^2}{(F^{-1}[K\phi])^{2(1-f)}}\right)^{-1/2}
\right]^{-{\frac{(3-m)}{4-m}}}, \label{pd2}
\end{equation}
where the constant $k_{1}$ is defined as $ k_{1}={\sqrt{3\pi}\over
4\mu^2}\left(  \frac{\mu^2C_{\phi}}{2 C_{\gamma}}\right)
^{{\frac{1}{4-m}}}(A\,f)^{{8-2m\over4-m}}%
\,(1-f)^{{\frac{m-3}{4-m}}} $.

Likewise,  the power spectrum as function of the number  $N$,
yields

\begin{equation}
{\mathcal{P}_{\mathcal{R}}}(N)=k_{2}\,\,(F[J(N)])^{\,\frac{1-m}{4-m}}
\left(J[N] \right)  ^{\frac{2f(4-m)+m-5}{4-m}} \left[1-
\epsilon\left(\alpha+ \frac{\beta
A^2f^2}{(J[N])^{2(1-f)}}\right)^{-1/2}
\right]^{-{\frac{(3-m)}{4-m}}},
\label{pd}%
\end{equation}
where  $J(N)$ and $k_{2}$ are given by $ J(N)=\left[{1+f(N-1)\over
Af} \right]^{\frac{1}{f}} $ and $ k_{2}=k_{1}K^{-\frac{1-m}{4-m}}
$, respectively.

The scalar spectral index $n_s$ given by $ n_s -1
=\frac{d \ln\,{\cal{P}_R}}{d \ln k}$,  where the wave number $k$, is
associated with the number of e-folds through  $d \ln k(\phi)=d
N(\phi)=(H/\dot{\phi})\,d\phi$. By using  Eqs. (\ref{at}) and
(\ref{pd}), this yields

\begin{equation}
n_{s}=1-{5-m-2f(4-m)\over Af(4-m)(F^{-1}[K\,\phi])^f} +n_2 +n_3,
\label{nss1}%
\end{equation}
where $n_2$ and $n_3$ are given by $$ n_{2}=\mu{1-m \over
4-m}\sqrt{{2(1-f)\over  Af}} {(F^{-1}[K\,\phi])^{-f/2}\over
\phi}\left[1- \epsilon\left(\alpha+ \frac{\beta
A^2f^2}{(F^{-1}[K\phi])^{2(1-f)}}\right)^{-1/2} \right]^{1/2}, $$
and

$$
n_3=\epsilon \frac{\beta Af(1-f)}{(F^{-1}[K\phi])^{2-f}}
\left(\alpha+ \frac{\beta
A^2f^2}{(F^{-1}[K\phi])^{2(1-f)}}\right)^{-3/2} \left[1-\epsilon
\left(\alpha+ \frac{\beta
A^2f^2}{(F^{-1}[K\phi])^{2(1-f)}}\right)^{-1/2} \right]^{-1},
$$
respectively.

Also,  $n_{s}$ can be written in terms of $N$. By using
Eqs.(\ref{N1}) and (\ref{nss1}), we obtain
\begin{equation}
n_{s}=1-\frac{5-m-2f(4-m)}{(4-m)[1+f(N-1)]}+n_{2N} +n_{3N}, \label{nswr}%
\end{equation}
where
$$
n_{2N}=\mu K{1-m \over 4-m}\sqrt{{2(1-f)\over  Af}}
{(J[N])^{-f/2}\over F[J(N)]}\left[1-\epsilon \left(\alpha+
\frac{\beta A^2f^2}{(J[N])^{2(1-f)}}\right)^{-1/2} \right]^{1/2},
$$
and
$$
n_{3N}=\epsilon \frac{\beta Af(1-f)}{(J[N])^{2-f}} \left(\alpha+
\frac{\beta A^2f^2}{(J[N])^{2(1-f)}}\right)^{-3/2}
\left[1-\epsilon \left(\alpha+ \frac{\beta
A^2f^2}{(J[N])^{2(1-f)}}\right)^{-1/2} \right]^{-1}.
$$
We observe  numerically from Eq. (\ref{nswr}) that the value of
$n_s \gg 1$, for the positive value of $\epsilon$, i.e., the
positive branch. This value of scalar spectral
index is disproved by the observational data. In this way,
the model of warm intermediate inflation on a warped DGP does not
work for the case $\epsilon=+1$. In the following, we will study
the negative branch in which $\epsilon=-1$.

\begin{figure}[th]
{\includegraphics[width=4.2in,angle=0,clip=true]{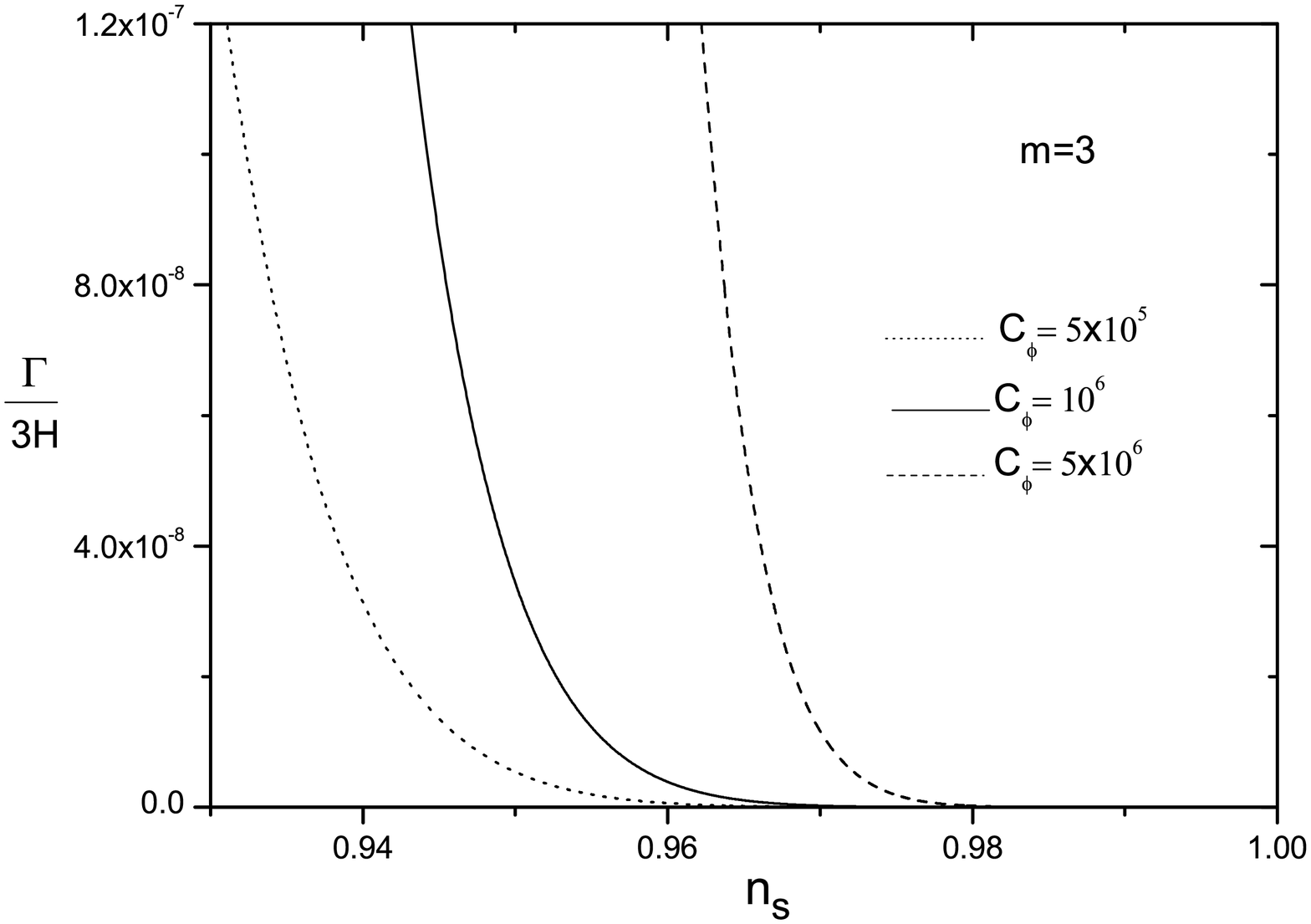}}

{{\hspace{-5.4
cm}\includegraphics[width=6.0in,angle=0,clip=true]{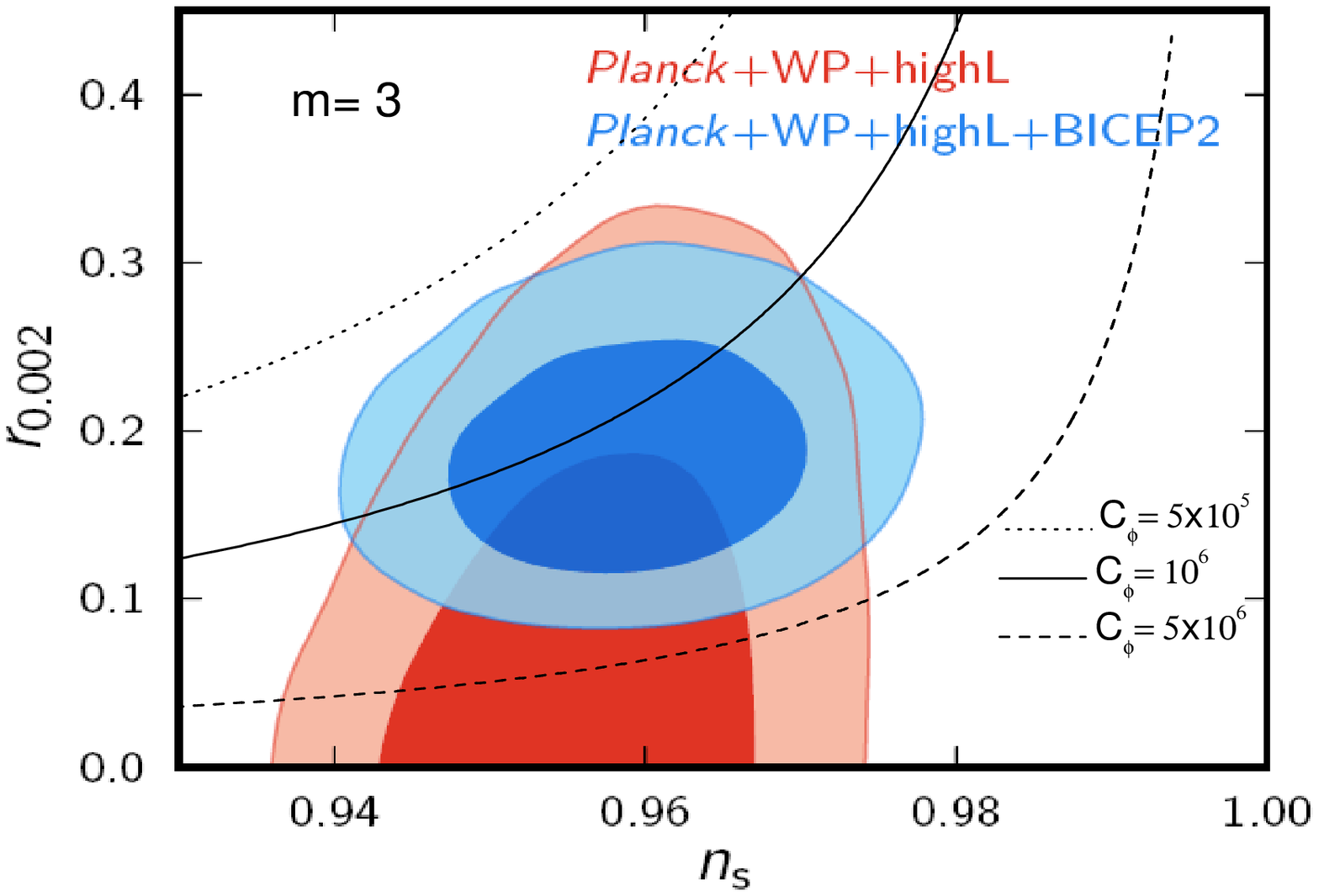}}}
{\vspace{-3 cm}\caption{The evolution of the ratio $\Gamma/3H$
versus the scalar spectrum index $n_s$ (upper panel) and the
evolution of the tensor-scalar ratio $r$ versus the scalar
spectrum index $n_s$ (lower panel) in the weak dissipative regime,
for three different values of the parameter $C_\phi$ and the
specific case $m=3$, i.e., $\Gamma\propto T^3/\phi^2$. In both
panels, the dotted, solid, and dashed lines correspond to the
pairs ($A=0.08$, $f=0.43$), ($A=0.19$, $f=0.42$) and ($A=0.23$,
$f=0.39$), respectively. Also, in these plots  we have taken the
values $C_\gamma=70$, $m_p=1$, $\mu=0.99$, $\eta=1$,
$\epsilon=-1$, ${\cal{A}}_0=1$ and $\Omega_{rc}=(\beta
H_0^2)^{-1}=0.014$\cite{Azizi:2011ys}.
 \label{figg1}}}
\end{figure}

\begin{figure}[th]
{{\hspace{-5.4
cm}\includegraphics[width=5.0in,angle=0,clip=true]{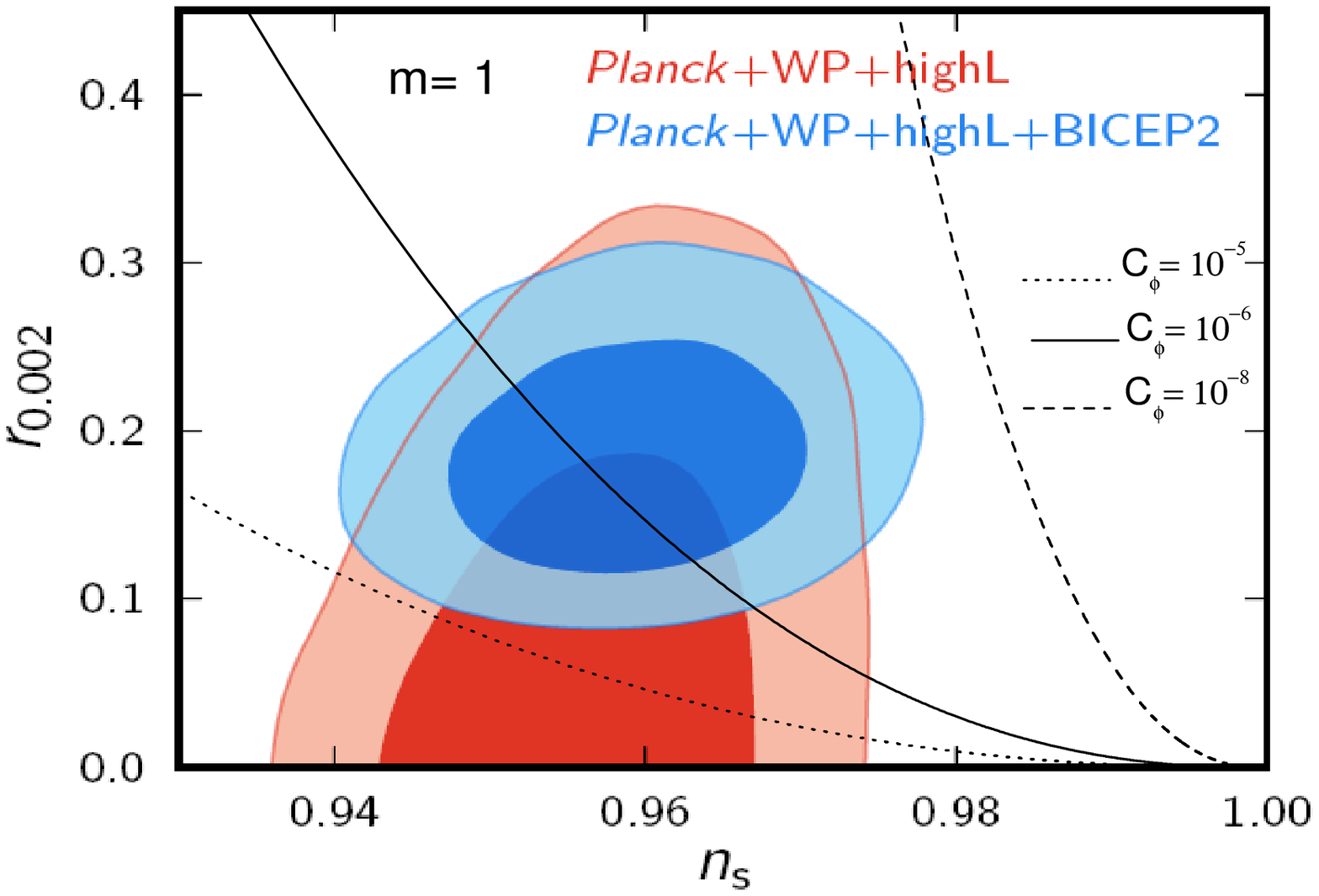}}}

{{\hspace{-5.4
cm}\includegraphics[width=5.0in,angle=0,clip=true]{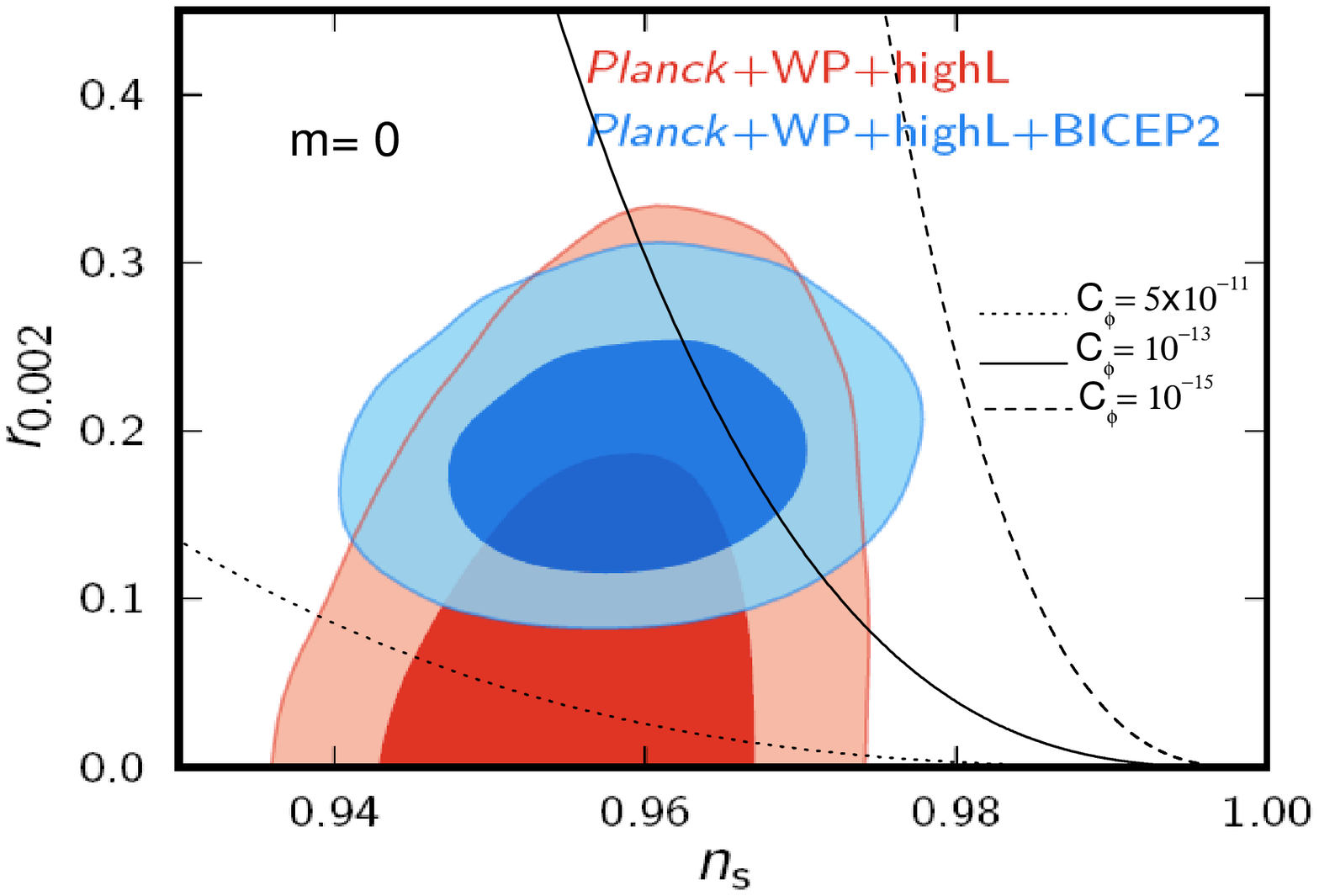}}}
{\vspace{-3 cm}\caption{ The upper and lower panels show the
evolution of the tensor-scalar ratio $r$ versus the scalar
spectrum index $n_s$ in the weak dissipative regime, for the cases
$m=1$ and $m=0$, respectively. In both panels we consider three
different values of the parameter $C_\phi$. For the case $m=1$
(upper panel), the dotted, solid, and dashed lines are for the
pairs ($A=0.67$, $f=0.29$), ($A=0.79$, $f=0.29$) and ($A=0.95$,
$f=0.29$). For the case $m=0$ (lower panel), the dotted, solid,
and dashed lines are for the pairs ($A=0.99$, $f=0.27$),
($A=1.38$, $f=0.27$) and ($A=1.75$, $f=0.26$). Also, in both
panels we have taken the values $C_\gamma=70$, $m_p=1$,
$\mu=0.99$, $\eta=1$, $\epsilon=-1$, ${\cal{A}}_0=1$ and
$\Omega_{rc}=(\beta H_0^2)^{-1}=0.014$\cite{Azizi:2011ys}.
 \label{figg2}}}
\end{figure}

On the other hand, it is well known that  the generation of tensor
perturbations during the inflationary epoch would produce
gravitational waves \cite{Bha}. In the warped DGP model, the amplitude
of gravitational waves \cite{Bou}, becomes
\begin{equation}
{\cal{P}}_g=\frac{64\pi}{m_{p}^{2}}(H/2\pi)^2\,G_{\gamma}^{2}(x),\label{ten}
\end{equation}
where $G_{\gamma}^{-2}(x)=\gamma+(1-\gamma)F(x)^{-2}$ is the
correction to standard general relativity. Here, the parameter
$\gamma=(\mu/m_p)^2$ and the function
$F(x)=[\sqrt{1+x}-x^2\sinh^{-1}(1/x)]^{-1/2}$, in which
$x=H/\bar{\mu}$, where $\bar{\mu}$ is the energy scale related
with the bulk curvature\cite{Bulk1}. In particular, in the case
when $x\rightarrow 0$, then $G_\gamma\rightarrow 1$ and thus reduces
to standard amplitude  of gravitational waves, where
${\cal{P}}_g=\frac{64\pi}{m_{p}^{2}}(H/2\pi)^2$. Also, in the case
when $\gamma\rightarrow 0$, the expression for ${\cal{P}}_g$,
coincides with the amplitude  of gravitational waves in the RS
case\cite{Bulk2}.

In this form, using Eqs.(\ref{pd2}) and (\ref{ten}), we may define
the tensor-to-scalar ratio as
$r=({\cal{P}}_g/\mathcal{P}_{\mathcal{R}})$,  and in terms of the
scalar field, this ratio, in the weak dissipative regime, can be
written as
\begin{equation}
r(\phi) \simeq\frac{16\,A^2\,f^2}{\pi\,
m_{p}^{2}(F^{-1}[K\phi])^{2(1-f)}}
\left[{G_{\gamma}^{2}(\phi)\over
{\cal{P}_{\cal{R}}}(\phi)}\right]. \label{Rk}
\end{equation}

Also, the tensor-to-scalar ratio can be rewritten in terms of the
number of e-folds $N$, as

\begin{equation}
r(N) \simeq\frac{16\,A^2\,f^2}{\pi\, m_{p}^{2}(J[N])^{2(1-f)}}
\left[{G_{\gamma}^{2}(N)\over {\cal{P}_{\cal{R}}}(N)}\right].
\label{Rk11}\end{equation}

In Fig.(\ref{figg1}) we show the evolution of  the ratio
$\Gamma/3H$ and the tensor-to-scalar ratio $r$  on the scalar
spectral index $n_s$ in the weak dissipative regime, for the
specific case $m=3$, in which the dissipation coefficient becomes
$\Gamma=C_\phi\,T^3/\phi^2$. In both panels we have considered
three different values of $C_\phi$. The upper panel shows the
dependence of $\Gamma/3H$ on the warm inflation and we confirm
that the model  remains in the weak dissipative regime
($\Gamma/3H<1$) during inflation. In the lower panel, we show the
two-dimensional marginalized constraints, at 68$\%$ and 95$\%$
levels of confidence, for the tensor-to-scalar ratio and the
scalar spectral index (taken BICEP2 experiment data in connection
with Planck + WP + highL, see Ref.\cite{B2}). From Eqs.(\ref{at}),
(\ref{G1}), (\ref{a7}) and (\ref{N1}) we can obtain in the weak
dissipative regime $R<1$, the ratio $\Gamma/3H$ as a function of
the number of e-folds $N$ i.e., $\Gamma/3H=f(N)$ and together with
the Eq.(\ref{nswr}),  we numerically obtain the parametric plot of
the curve $\Gamma/3H=\Gamma/3H(n_s)$ (upper panel). Analogously,
we consider Eqs.(\ref{nswr}) and (\ref{Rk11}) and we numerically
find the parametric plot of the consistency relation $r=r(n_s)$
(lower panel). Here, we consider that the range for the number of
e-folds $N$ is given by $30\leq N\leq 120$. In these plots we use
 the
values of $C_\gamma=70$, $m_p=1$, $\mu=0.99$, $\eta=1$,
$\epsilon=-1$, ${\cal{A}}_0=1$ and $\Omega_{rc}=(\beta
H_0^2)^{-1}=0.014$\cite{Azizi:2011ys}, in which $H_0=68.6$ km
s$^{-1}$ Mpc$^{-1}$. In addition, we numerically make use of
Eqs.(\ref{pd}) and (\ref{nswr}), and we find that $A=0.08\times
10^{-6}$ and $f=0.43$ for the value of $C_\phi=5\times10^{5}$ for
which $n_s=0.96$,
 ${\cal{P}_{\cal{R}}}=2.43\times10^{-9}$ and $N=60$. Similarly,
 $C_\phi=10^{6}$ corresponds to $A=0.19$, $f=0.42$; for the
case in which $C_\phi=5 \times10^{6}$, then $A=0.23$, $f=0.39$.
 From the lower plot we find  that the range for the parameter
$C_\phi$, in the special case $m=3$, is given by $5\times
10^5<C_\phi<5\times10^{6}$, which is well corroborated from the
BICEP2 experiment and also from Planck data.

In Fig.(\ref{figg2}) we show the evolution of the tensor to scalar
ratio $r$ on the scalar spectral index $n_s$ for the weak
dissipative regime, where in the upper panel we fix $m=1$ and in
the lower panel $m=0$. As before we consider three values of
$C_\phi$. Again, we use Eqs.(\ref{nswr}) and (\ref{Rk11}) and we
numerically find the parametric plot $r=r(n_s)$ for $m=1$ and
$m=0$,
 where $C_\gamma=70$, $m_p=1$, $\mu=0.99$, $\eta=1$,
$\epsilon=-1$, ${\cal{A}}_0=1$ and $\Omega_{rc}=(\beta
H_0^2)^{-1}=0.014$\cite{Azizi:2011ys}. As before, we solve
Eqs.(\ref{pd}) and (\ref{nswr}), and we find that for the case
$m=1$ (upper panel), the dotted, solid, and dashed lines
correspond to the pairs ($A=0.67$, $f=0.29$), ($A=0.79$, $f=0.29$)
and ($A=0.95$, $f=0.29$), respectively. For the case $m=0$ (lower
panel), the dotted, solid, and dashed lines correspond to the
pairs ($A=0.99$, $f=0.27$), ($A=1.38$, $f=0.27$) and ($A=1.75$,
$f=0.26$), respectively. From the upper plot we obtain  the range
for $C_\phi$ in the specific case $m=1$, given by
$10^{-8}<C_\phi<10^{-5}$. From the lower plot we observe  that the
range for $C_\phi$ in the  case $m=0$ is given by $
10^{-15}<C_\phi<5\times10^{-11}$.  These ranges of the parameter
$C_\phi$, for both models, are well supported by BICEP2 experiment
data and Planck data. Finally, for the special case $m=-1$, i.e.,
$\Gamma\propto \phi^2/T^{1}$, we find that the range for $C_\phi$
becomes $ 10^{-22}<C_\phi<10^{-16}$, where $C_\phi=10^{-22 }$
corresponds to $A=2.22$, $f=0.25$ and for the case in which
$C_\phi=10^{-16}$, then $A=1.27$, $f=0.24$ (not shown). In this
form,  we observe that when we decrease the value of the parameter
$m$ the range of the parameter $C_\phi$ also decreases.  Also, we
noted that in the weak dissipative regime the constraint on the
parameter $C_\phi$ only arises from the BICEP2 experiment (or
Planck data) and not from of the ratio $\Gamma/3H$.

\begin{figure}[th]
{{\hspace{.4
cm}\includegraphics[width=4.0in,angle=0,clip=true]{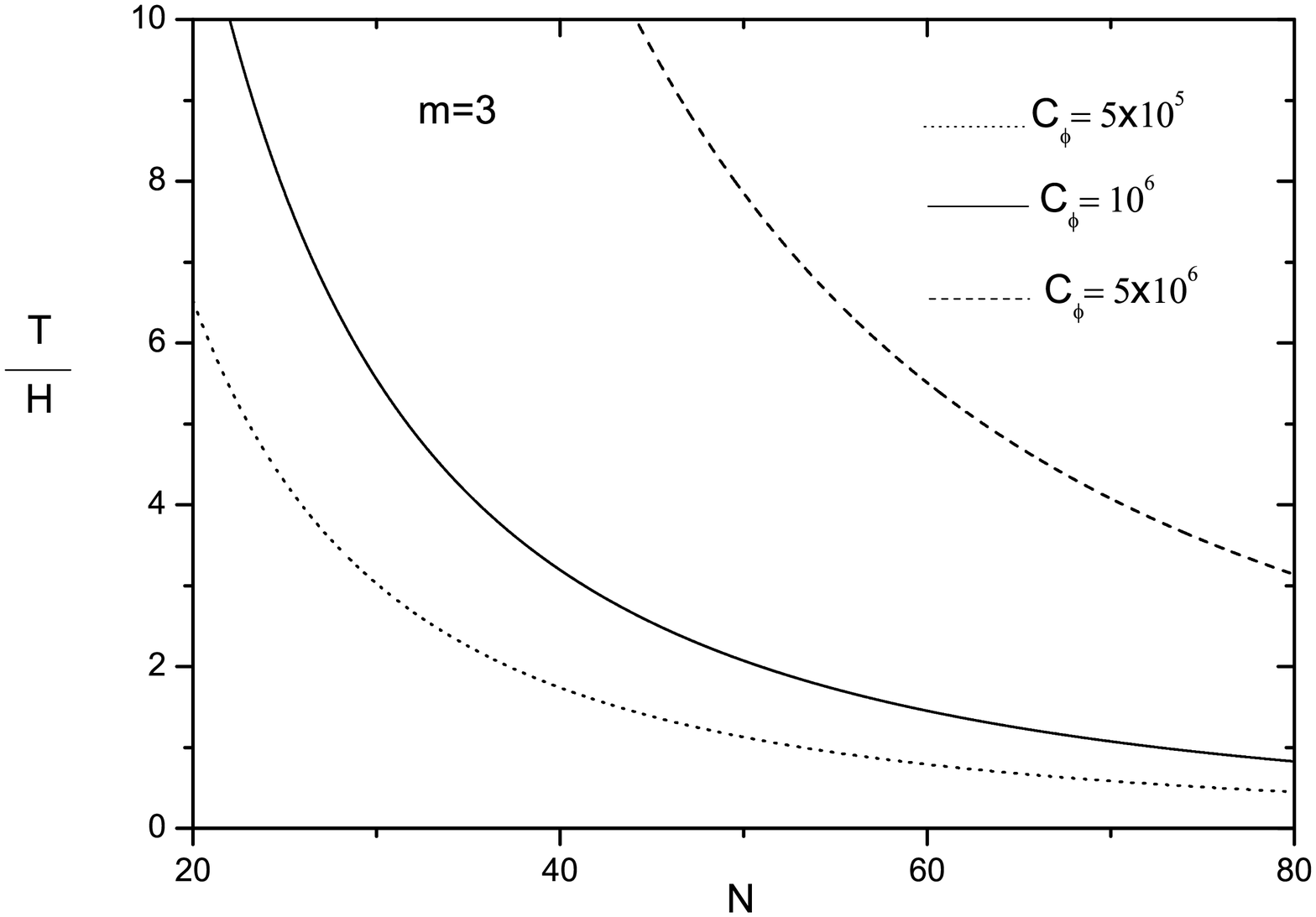}}}

{{\hspace{0.4
cm}\includegraphics[width=4.0in,angle=0,clip=true]{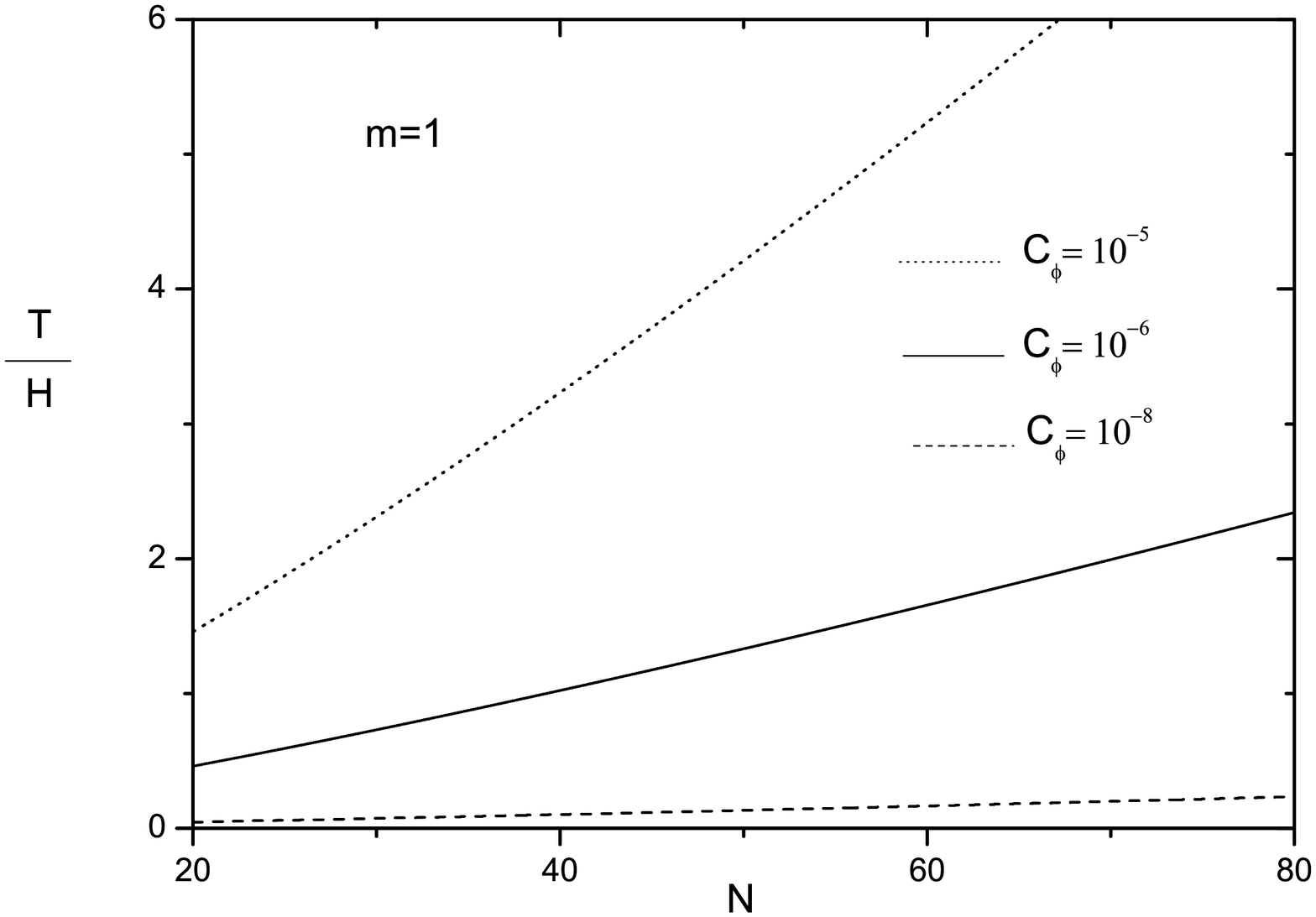}}}
{\vspace{-1 cm}\caption{ The upper and lower panels show the
evolution of the ratio $T/H$ versus the number of e-folds  $N$ in
the weak dissipative regime, for the cases $m=3$ and $m=1$,
respectively. Here, we have used  the same parameters of the
Figs.(1) and (2).
 \label{THD}}}
\end{figure}

In Fig.(\ref{THD}) we show the evolution of the ratio $T/H$ on the
number of e-folds  $N$ for the weak dissipative regime, where in
the upper panel we fix $m=3$ and in the lower panel $m=1$. In
order to write down values for the ratio $T/H$  and the number of
e-folds $N$, for the special cases $m=3$ and $m=1$,  we utilize
Eqs. (\ref{at}), (\ref{TD}) , (\ref{a7}) and (\ref{N1}), together
with the same parameters from Figs.(1) and (2). From  the upper
panel we note that the value $C_\phi>5\times 10^{5} $ is well
supported by the necessary condition for the warm inflation
scenario i.e., $T>H$. In particular, for the value $C_\phi=10^6$
and evaluating for the value of $N=60$ (where $n_s=0.96$), we get
that the value of the ratio $T/H\simeq 1.47$, for the value
$C_\phi=5\times10^6$ corresponds to $T/H\simeq 5.51$ and for
$C_\phi=5\times10^5$ it corresponds to $T/H\simeq 0.79.$ In this
way,  we find that the range for the parameter $C_\phi$, in the
special case $m=3$, is given by $5\times
10^5<C_\phi<5\times10^{6}$, which are well corroborated from
Planck data and the BICEP2 experiment together with the condition
for warm inflation $T/H>1$.

From the lower panel we observe that the value
$C_\phi\gtrsim10^{-6}$ is well supported by the condition for warm
inflation i.e., $T/H>1$ ( for $N\geq 60$). In particular, for the
value $C_\phi=10^{-5}$, we obtain that $T/H\simeq 5.28$ for $N=60$
and for the value $C_\phi=10^{-6}$ corresponds to $T/H\simeq
1.67$. For the values  $C_\phi=10^{-8}$ and $N=60$ corresponds to
$T/H\simeq 0.17$ and the model of the weak dissipative regime is
disfavored from the essential condition for the warm inflation
scenario, since the ratio  $T/H<1$. It interesting to note that
from the condition $T/H>1$, we have found a lower bound for the
parameter $C_\phi$.  In this form,  for the value $m=1$ we can set
a new constraint for the parameter $C_\phi$, given by
$10^{-6}\lesssim C_\phi <10^{-5}$ from BICEP2 experiment (or
Planck data) and the condition  $T/H>1$.

For the case $m=0$ the evolution of the ratio $T/H$ on the number
of e-folds  $N$ for the weak dissipative regime, is similar to the
case $m=1$ (figure not shown).  As before, we note that the value
$C_\phi\gtrsim10^{-13}$  is well supported by the condition for
the warm inflation scenario $T/H>1$. Also, we find that that for
the values $C_\phi=10^{-15}$ and $N=60$  corresponds to $T/H\simeq
0.15$ and then $T/H<1$.  Again, we observe that from the condition
$T/H>1$ we have obtained a lower bound for $C_\phi$, and the range
for this parameter
 from BICEP2 experiment (or
Planck data)
 and $T/H>1$, is given by
$10^{-6}\lesssim C_\phi <10^{-5}$. Finally, for the special  case
$m=-1$,  we note that the value $C_\phi\gtrsim10^{-19}$ is well
supported by the condition $T/H>1$ for $N\geq 60$(not shown).
 Also, in particular  we note that for
the values  $C_\phi=10^{-22}$  and $N=60$ the model of the weak
dissipative regime is disfavored from the  condition for the warm
inflation scenario, since the rate $T/H\simeq 0.14$. Again, we
observe that from  $T/H>1$ we have found a lower bound for
$C_\phi$. In this form, for $m=-1$ the new constraint for the
parameter $C_\phi$ from the Planck data (or BICEP2)
 and $T/H>1$, is given by
$10^{-19}\lesssim C_\phi <10^{-16}$.  Finally, we note that in the
weak dissipative regime from the essential condition for warm
inflation $T/H>1$, we have found a lower bound for the parameter
$C_\phi$ and this bound becomes independent from Planck data and
BICEP2 experiment.

\subsection{ The strong dissipative regime.\label{4}}

Now we study the case of the strong dissipative regime, i.e.,
$\Gamma>3H$. By using  Eqs.(\ref{a4}) and (\ref{G1}), and
considering  the intermediate expansion given by Eq.(\ref{at}), we
obtain a relation between the scalar field and cosmological
time. However, we must study two cases separately, namely
$m=3$, and $m\neq3$. For the case $m=3$, the solution for the
scalar field yields
\begin{equation}
\phi(t)-\phi_0=\exp\left[\frac{\tilde{F}[t]}{\tilde{K}}\right],\label{phis}
\end{equation}
again $\phi(t=0)=\phi_0$ is an  integration  constant,  the
constant $\tilde{K}$ is defined by
$$
\tilde{K}\equiv
\left(\frac{C_{\phi}}{6\mu^2}\right)^\frac{1}{2}\left(\frac{3\mu^2}{2C_{\gamma}}\right)^\frac{3}{8}\left(1-f\right)
^\frac{7}{8}\beta^{-\frac{\left(5f+2\right)}{16\left(1-f\right)}}
\left(Af\right)^{-\frac{7}{8\left(1-f\right)}},
$$
and the new function $\tilde{F}[t]$, corresponds to the incomplete
Lauricella function\cite{libro}, given by
$$
\tilde{F}[t]\equiv\frac{\left[\epsilon\left(\alpha+\frac{\beta}{t^{2(1-f)}
}\right)^{-1/2}\right]^{2\left(\nu-1\right)}}{2\left(\nu-1\right)}
F^{(3)}_{D}[2\left(\nu-1\right);\nu,\nu,-\frac{1}{8}
,2\nu-1,\sqrt{\alpha},-\sqrt{\alpha},
\epsilon\left(\alpha+\frac{\beta}{t^{2(1-f)} }\right)^{-1/2} ],
$$
where the constant $\nu$ is defined as
$\nu=\frac{18-11f}{16(1-f)}$.

 For the  case $m\neq 3$, the solution for the scalar field can be written as
\begin{equation}
\varphi(t)-\varphi_0=\frac{\tilde{F}_{m}[t]}{\tilde{K}_{m}},\label{phism}
\end{equation}
where now  the new scalar field $\varphi$, is defined as
$\varphi(t)=\frac{2}{3-m}{\phi(t)}^\frac{3-m}{2}$, the constant
$\tilde{K}_{m}=\left(\frac{C_{\phi}}{6\mu^2}\right)^\frac{1}{2}\left(\frac{3\mu^2}{2C_{\gamma}}\right)^\frac{m}{8}
\left(1-f\right)^\frac{4+m}{8}\beta^\frac{4+f\left(m-8\right)-2m}{16\left(1-f\right)}
\left(Af\right)^{-\frac{4+m}{8\left(1-f\right)}}$, and the
function $\tilde{F}_{m}[t]$, is defined as
$$
\tilde{F}_{m}[t]\equiv\frac{\left[\epsilon\left(\alpha+\frac{\beta}{t^{2(1-f)}
}\right)^{-1/2}\right]^{2\left(\nu-1\right)}}{2\left(\nu-1\right)}
F^{(3)}_{D}[2\left(\nu-1\right);\nu,\nu,-\frac{(4-m)}{8}
,2\nu-1,\sqrt{\alpha},-\sqrt{\alpha},
\epsilon\left(\alpha+\frac{\beta}{t^{2(1-f)} }\right)^{-1/2} ],
$$ in which  the constant $\nu_{m}=\frac{2(6+m)-f(8+m)}{16(1-f)}.
$

In this regime, the Hubble parameter as a function of the inflaton
field $\phi$, for both cases  becomes

\begin{equation}
H(\phi)=\frac{Af}{(\tilde{F}^{-1}[\tilde{K}\ln\phi])^{1-f}},\,\,\,\,\,\,\,\,\mbox{for
$m=3$},\label{H3}
\end{equation}
 and

\begin{equation}
H(\phi)=\frac{Af}{(\tilde{F}_{m}^{-1}[\tilde{K}_{m}\varphi])^{1-f}},\,\,\,\,\,\,\,\,\mbox{for
$m\neq3$}.\label{Hm}
\end{equation}

As before, considering  the slow-roll approximation,  the scalar
potential in the strong dissipative regime from Eq.(\ref{a5}),
yields
$$
V(\phi)\simeq\frac{\eta\rho_0 }{2}\left(\alpha+ \frac{\beta A^2f^2}{(\tilde{F}^{-1}[\tilde{K}\ln\phi])^{2(1-f)}}\right)
\left[1-
\epsilon\left(\alpha+ \frac{\beta A^2f^2}{(\tilde{F}^{-1}[\tilde{K}\ln\phi])^{2(1-f)}}\right)^{-1/2}
\right]^{2}-\frac{{\cal{A}}_0 ^{2}\rho_0}{2\eta}+
$$
\begin{equation}
-\frac{3}{2}{\mu^2\,\frac{Af(1-f)}{(\tilde{F}^{-1}[\tilde{K}\ln\phi])^{2-f}}}\,\left[1-
\epsilon\left(\alpha+ \frac{\beta A^2f^2}{(\tilde{F}^{-1}[\tilde{K}\ln\phi])^{2(1-f)}}\right)^{-1/2}
\right]
,\label{VV3}
\end{equation}
for the case in which m=3, and

$$
V(\phi)\simeq\frac{\eta\rho_0 }{2}\left(\alpha+ \frac{\beta A^2f^2}{(\tilde{F}_{m}^{-1}[\tilde{K}_{m}\varphi])^{2(1-f)}}\right)
\left[1-
\epsilon\left(\alpha+ \frac{\beta A^2f^2}{(\tilde{F}_{m}^{-1}[\tilde{K}_{m}\varphi])^{2(1-f)}}\right)^{-1/2}
\right]^{2}-\frac{{\cal{A}}_0 ^{2}\rho_0}{2\eta}+
$$
\begin{equation}
-\frac{3}{2}{\mu^2\,\frac{Af(1-f)}{(\tilde{F}_{m}^{-1}[\tilde{K}_{m}\varphi])^{2-f}}}\,\left[1-
\epsilon\left(\alpha+ \frac{\beta A^2f^2}{(\tilde{F}_{m}^{-1}[\tilde{K}_{m}\varphi])^{2(1-f)}}\right)^{-1/2}
\right]
,\label{VVm}
\end{equation}
for $m\neq3.$

Also, the dissipation coefficient from
Eqs.(\ref{G1}),(\ref{phis}), and (\ref{phism}) becomes
\begin{equation}
\Gamma (\phi)=\delta \phi^{-2}
(\tilde{F}^{-1}[\tilde{K}\ln\phi])^{-\frac{3(2-f)}{4}}
\left[1-\epsilon \left(\alpha+\beta
H^{2}\right)^{-1/2}\right]^{3/4},\,\,\,\,\,\mbox{for
$m=3$},\label{gammas3}
\end{equation}
where the constant $\delta=C_{\phi} \left(\frac{3\mu^{2}Af(1-f)}{2
C_{\gamma}}\right)^{3/4}$, and
\begin{equation}
\Gamma (\phi)=\delta_{m} \phi^{1-m}
(\tilde{F}_{m}^{-1}[\tilde{K}_{m}\varphi])^{-\frac{m(2-f)}{4}}
\left[1-\epsilon \left(\alpha+\beta
H^{2}\right)^{-1/2}\right]^{m/4},\,\,\,\,\,\mbox{for
$m\neq3$}.\label{gammasm}
\end{equation}
Here the constant $\delta_{m}$ is defined as $\delta_{m}=C_{\phi}
\left(\frac{3\mu^{2}Af(1-f)}{2 C_{\gamma}}\right)^{m/4}$.

Now for this regime, the dimensionless slow-roll parameter
$\varepsilon$ is given by
$\varepsilon=-\frac{\dot{H}}{H^2}=\frac{1-f}{Af(\tilde{F}^{-1}[\tilde{K}\ln\phi])^f}$,
for the case in which $m=3$; for the case $m\neq3$, we get
$\varepsilon=\frac{1-f}{Af(\tilde{F}_{m}^{-1}[\tilde{K}_{m}\varphi])^f}$
.

As before, the condition for inflation $\ddot{a}>0$ is satisfied
when the scalar field
$\phi>\exp\left[\frac{1}{\tilde{K}}\tilde{F}\left[\left(\frac{1-f}{Af}\right)^{1/f}\right]\right]$,
for the case in which $m=3$; for $m\neq3$, the
condition is satisfied for the new scalar field  when
$\varphi>\frac{1}{\tilde{K}_{m}}\tilde{F}_{m}\left[\left(\frac{1-f}{Af}\right)^{1/f}\right]$.
As in the weak dissipative regime, the inflationary scenario
begins at the earliest possible scenario in which $\varepsilon=1$.
Here,
$\phi_{1}=\exp\left[\frac{1}{\tilde{K}}\tilde{F}\left[\left(\frac{1-f}{Af}\right)^{1/f}\right]\right]$,
for the case in which $m=3$, and
$\varphi_{1}=\frac{1}{\tilde{K}_{m}}\tilde{F}_{m}\left[\left(\frac{1-f}{Af}\right)^{1/f}\right]$,
for the case $m\neq3$.

In the strong dissipative regime, the expression for the number of
e-folds between two different values  $\phi_1$ and $\phi_2$, from
Eqs.(\ref{at}), (\ref{phis}), and(\ref{phism}), becomes
\begin{equation}
N=\int_{t_1}^{t_{2}}\,H\,dt=A\,\left[(\tilde{F}^{-1}[\tilde{K}\ln\phi_{2}])^{f}-(\tilde{F}^{-1}[\tilde{K}\ln\phi_{1}])^{f}\right],\,\,\,\,\mbox{for
$m=3$},\label{Ns3}
\end{equation}
and
\begin{equation}
N=A\,\left[(\tilde{F}_{m}^{-1}[\tilde{K}_{m}\varphi_{2}])^{f}-(\tilde{F}_{m}^{-1}[\tilde{K}_{m}\varphi_{1}])^{f}\right],\,\,\,\,\mbox{for
$m\neq3$}.\label{Nsm}
\end{equation}

On the other hand, as before the power spectrum related to the curvature
spectrum could be written as
$\cal{P}_{\cal{R}}\simeq\frac{H}{\dot{\phi}}\delta\phi$, where in
the strong dissipative regime ,i.e.,$\Gamma>3H$, we get that
$\delta\phi^2\simeq \frac{k_{F}T}{2\pi^2}$ \cite{warm}, in which
$k_{F}=\sqrt{\Gamma H}$. In this way, from Eqs.(\ref{at}),
(\ref{rh-1}), and (\ref{G1}), the expression for the spectrum of
the scalar perturbation yields
\begin{equation}
P_{\mathcal{R}}\simeq\frac{H^{\frac{5}{2}}\Gamma^{\frac{1}{2}}T}{2\pi^{2}%
\dot{\phi}^{2}}=\frac{C_{\phi}^{3/2}}{2\pi^{2}}\left(\frac{1}{6\mu^{2}}\right)
\left(\frac{3\mu^{2}}{2C_{\gamma}}\right)^{\frac{3m+2}{8}}\phi^{\frac{3(1-m)}{2}}H^{\frac{3}{2}}
\left(-\dot{H}\right)^{\frac{3m-6}{8}}\left[1-\epsilon\left(\alpha+\beta H^{2}
\right)^{-1/2}\right]^{\frac{3m-6}{8}}.\label{PRRm}%
\end{equation}

As before, it is necessary to separate the specific  cases $m=3$
and $m\neq3$. Replacing Eqs.(\ref{at}), (\ref{phis}), and
(\ref{phism}) in Eq.(\ref{PRRm}), we can obtain  the power
spectrum  in terms of the scalar field for both values of $m$. In
this form, for the case in which $m=3$, we get
\begin{equation}
{\cal{P}_{\cal{R}}}=k\phi^{-3}(\tilde{F}^{-1}[\tilde{K}\ln\phi])^{\frac{3(5f-6)}{8}}
\left[1-\epsilon\left(\alpha+\beta A^{2} f^{2}(\tilde{F}^{-1}
[\tilde{K}\ln\phi])^{-2(1-f)}\right)^{-1/2}\right]^{\frac{3}{8}},\label{Prs3}
\end{equation}
where the constant $k$ is defined as
$k=\frac{C_{\phi}^{3/2}}{2\pi^{2}}\frac{1}{6\mu^2}\left(\frac{3\mu^2}{2C_{\gamma}}\right)^{\frac{11}{8}}
\left(1-f\right)^{\frac{3}{8}}\left(Af\right)^{\frac{15}{8}}.$
For the specific case in which $m\neq3$, the spectrum of the
scalar perturbation yields
\begin{equation}
{\cal{P}_{\cal{R}}}=k_{m}\phi^{\frac{3(1-m)}{2}}(\tilde{F}_{m}^{-1}[\tilde{K}_{m}\varphi])^{\frac{3[f(m+2)-2m]}
{8}}\left[1-\epsilon\left(\alpha+\beta A^{2}
f^{2}(\tilde{F}_{m}^{-1}[\tilde{K}_{m}\varphi])^{-2(1-f)}\right)^{-1/2}\right]^{\frac{3m-6}{8}},\label{Prsm}
\end{equation}
where the constant
$k_{m}=\frac{C_{\phi}^{3/2}}{2\pi^{2}}\frac{1}{6\mu^2}\left(\frac{3\mu^2}{2C_{\gamma}}\right)^{\frac{3m+2}{8}}
\left(1-f\right)^{\frac{3m-6}{8}}\left(Af\right)^{\frac{3m+6}{8}}.$

In order to manipulate numerically the equations, it is useful to
rewrite the scalar power spectrum in terms of the number of
e-folds. By using Eqs.(\ref{Ns3}), and (\ref{Nsm}), the above
expressions becomes
\begin{equation}
{\cal{P}_{\cal{R}}}=k\exp\left(-\frac{3}{\tilde{K}}\tilde{F}\left[J[N]\right]\right)(J[N])^{\frac{3(5f-6)}{8}}
\left[1-\epsilon\left(\alpha+\beta A^{2} f^{2}(J[N])^{-2(1-f)}\right)^{-1/2}\right]^{\frac{3}{8}},\label{Prs3N}
\end{equation}
for the case in which $m=3$, and
\begin{equation}
{\cal{P}_{\cal{R}}}=\tilde{\gamma}_{m} (\tilde{F}_{m}[J[N]])^{\frac{3(1-m)}{3-m}}
 (J[N])^{\frac{3[f(m+2)-2m]}{8}}\left[1-\epsilon\left(\alpha+\beta A^{2}
f^{2}(J[N])^{-2(1-f)}\right)^{-1/2}\right]^{\frac{3m-6}{8}},\label{PrsmN}
\end{equation}
for the case $m\neq 3$. Here, the constant $\tilde{\gamma}_{m}$ is
given by
$\tilde{\gamma}_{m}=k_{m}\left(\frac{2\tilde{K}_{m}}{3-m}\right)^{-\frac{3(1-m)}{3-m}}$.

From Eqs. (\ref{Prs3}) and (\ref{Prsm}), the scalar spectral index
$n_s$ in the strong dissipative regime yields
\begin{equation}
n_s\simeq1+\frac{3(5f-6)}{8Af}(\tilde{F}^{-1}[\tilde{K}\ln\phi])^{-f}+n_{\epsilon1}+n_{\epsilon2},\label{ns3}
\end{equation}
where $n_{\epsilon1}$ and $n_{\epsilon2}$, are given by
$$n_{\epsilon1}=-3\tilde{K}(1-f)^{1/8}\left(Af\right)^{-3/8}(\tilde{F}^{-1}[\tilde{K}\ln\phi])^{\frac{2-3f}{8}}
$$
$$\times\left[1-\epsilon\left(\alpha+\beta A^{2}f^{2}(\tilde{F}^{-1}[\tilde{K}\ln\phi])^{-2(1-f)}\right)^{-1/2}\right]^{1/8},
$$
and $$n_{\epsilon2}=-\epsilon \frac{3}{8}\beta Af(1-f)(\tilde{F}^{-1}[\tilde{K}\ln\phi])^{f-2}\left(\alpha+\beta A^{2}f^{2}(\tilde{F}^{-1}
[\tilde{K}\ln\phi])^{-2(1-f)}\right)^{-3/2}
$$
$$\times\left[1-\epsilon\left(\alpha+\beta A^{2}f^{2}(\tilde{F}^{-1}[\tilde{K}\ln\phi])^{-2(1-f)}\right)^{-1/2}\right]^{-1},
$$ for the specific case in which $m=3$. For the case  $m\neq3$ we have
\begin{equation}
n_s\simeq1+\frac{3[f(2+m)-2m]}{8Af}(\tilde{F}_{m}^{-1}[\tilde{K}_{m}\varphi])^{-f}+n_{\epsilon1_m}+n_{\epsilon2_m},\label{nsm}
\end{equation}
where $$n_{\epsilon1_m}=\frac{3(1-m)}{2}\tilde{K}_{m}(1-f)^{\frac{4-m}{8}}\left(Af\right)^{-\frac{m}{8}}\phi^{\frac{m-3}{2}}(\tilde{F}_{m}^{-1}
[\tilde{K}_{m}\varphi])^{\frac{[m(2-f)-4]}{8}}
$$
$$\times\left[1-\epsilon\left(\alpha+\beta A^{2}f^{2}(\tilde{F}_{m}^{-1}[\tilde{K}_{m}\varphi])^{-2(1-f)}\right)^{-1/2}\right]^{\frac{4-m}{8}},
$$
and $$n_{\epsilon2_m}=-\epsilon \frac{(3m-6)}{8}\beta Af(1-f)(\tilde{F}_{m}^{-1}[\tilde{K}_{m}\varphi])^{f-2}\left(\alpha+\beta A^{2}f^{2}
(\tilde{F}_{m}^{-1}[\tilde{K}_{m}\varphi])^{-2(1-f)}\right)^{-3/2}
$$
$$\times\left[1-\epsilon\left(\alpha+\beta A^{2}f^{2}(\tilde{F}_{m}^{-1}[\tilde{K}_{m}\varphi])^{-2(1-f)}\right)^{-1/2}\right]^{-1}.
$$

Analogously as before,  the scalar spectral index $n_s$ can be
rewritten in terms of the number of e-folds. By considering
Eqs.(\ref{Ns3}), and (\ref{Nsm}), the scalar spectral index
results
\begin{equation}
n_s\simeq1+\frac{3(5f-6)}{8Af}(J[N])^{-f}+n_{\epsilon1}[J[N]]+n_{\epsilon2}[J[N]],\label{ns3N}
\end{equation}
where $$n_{\epsilon1}[J[N]]=-3\tilde{K}(1-f)^{1/8}\left(Af\right)^{-3/8}(J[N])^{\frac{2-3f}{8}}
$$
$$\times\left[1-\epsilon\left(\alpha+\beta A^{2}f^{2}(J[N])^{-2(1-f)}\right)^{-1/2}\right]^{1/8},
$$
and $$n_{\epsilon2}[J[N]]=-\epsilon \frac{3}{8}\beta Af(1-f)(J[N])^{f-2}\left(\alpha+\beta A^{2}f^{2}(J[N])^{-2(1-f)}\right)^{-3/2}
$$
$$\times\left[1-\epsilon\left(\alpha+\beta A^{2}f^{2}(J[N])^{-2(1-f)}\right)^{-1/2}\right]^{-1},
$$ for the specific case in which $m=3$. For the case $m\neq3$ the spectral index
can be written as
\begin{equation}
n_s\simeq1+\frac{3[f(2+m)-2m]}{8Af}(J[N])^{-f}+n_{\epsilon1_m}[J[N]]+n_{\epsilon2_m}J[[N]],\label{nsmN}
\end{equation}
where $$n_{\epsilon1_m}[J[N]]=\frac{3(1-m)}{3-m}(1-f)\left(Af\right)^{\frac{[(f-2)m-4]}{8(1-f)}}\beta^{\frac{[4+f(m-8)-2m]}{16(1-f)}}
(\tilde{F}_{m}[J[N]])^{-1}(J[N])^{\frac{[m(2-f)-4]}{8}}
$$
$$\times\left[1-\epsilon\left(\alpha+\beta A^{2}f^{2}(J[N])^{-2(1-f)}\right)^{-1/2}\right]^{\frac{4-m}{8}},
$$
and $$n_{\epsilon2_m}[J[N]]=-\epsilon \frac{(3m-6)}{8}\beta Af(1-f)(J[N])^{f-2}\left(\alpha+\beta A^{2}f^{2}(J[N])^{-2(1-f)}\right)^{-3/2}
$$
$$\times\left[1-\epsilon\left(\alpha+\beta A^{2}f^{2}(J[N])^{-2(1-f)}\right)^{-1/2}\right]^{-1}.
$$

On the other hand, from  Eqs.(\ref{ten}) and  (\ref{Prsm}),
 the tensor-to scalar ratio, for the warped DGP model
in the strong dissipative regime, for the case $m=3$, can be
written  as

\begin{equation}
r=\zeta\,\phi^{3}(\tilde{F}^{-1}[\tilde{K}\ln\phi])^{\frac{f+2}{8}}\left[1-\epsilon\left(\alpha+\beta A^{2} f^{2}(\tilde{F}^{-1}
[\tilde{K}\ln\phi])^{-2(1-f)}\right)^{-1/2}\right]^{-\frac{3}{8}},\label{rfi3}
\end{equation}
where the constant
$\zeta=24\left(\frac{\mu^2}{m_p^2}\right)\frac{(Af)^{\frac{1}{8}}}{C_{\phi}^{3/2}}
\left(\frac{2C_{\gamma}}{3\mu^{2}}\right)^{\frac{11}{8}}(1-f)^{-\frac{3}{8}}$,
and for the specific case in which $m\neq3$, we get
\begin{equation}
r=\zeta_{m}\,\phi^{-\frac{3}{2}(1-m)}(\tilde{F}_{m}^{-1}[\tilde{K}_{m}\varphi])^{\frac{1}{8}[-16+f(10-3m)+6m]}\left[1-\epsilon
\left(\alpha+\beta A^{2} f^{2}(\tilde{F}_{m}^{-1}[\tilde{K}_{m}\varphi])^{-2(1-f)}\right)^{-1/2}\right]^{-\frac{1}{8}(3m-6)},\label{rfim}
\end{equation}
where $\zeta_m=24\left(\frac{\mu^2}{m_p^2}\right)\frac{(Af)^{\frac{1}{8}(10-3m)}}{C_{\phi}^{3/2}}
\left(\frac{2C_{\gamma}}{3\mu^{2}}\right)^{\frac{3m+2}{8}}(1-f)^{-\frac{(3m-6)}{8}}$.

Analogously as before,  the tensor-to-scalar ratio $r$ as a
function of the number e-folds $N$, becomes
\begin{equation}
r(N)=\zeta\,\exp\left(\frac{3}{\tilde{K}}\tilde{F}\left[J[N]\right]\right)(J[N])^{\frac{f+2}{8}}\left[1-\epsilon\left(\alpha+
\beta A^{2} f^{2}(J[N])^{-2(1-f)}\right)^{-1/2}\right]^{-\frac{3}{8}},\label{rfi3N}
\end{equation}
for the case in which $m=3$, and
\begin{equation}
r=\tilde{\zeta}_{m} (\tilde{F}_{m}[J[N]])^{-\frac{3(1-m)}{3-m}}(J[N])^{\frac{1}{8}[-16+f(10-3m)+6m]}\left[1-\epsilon
\left(\alpha+\beta A^{2} f^{2}(J[N])^{-2(1-f)}\right)^{-1/2}\right]^{-\frac{1}{8}(3m-6)},\label{rfimN}
\end{equation}
for the case $m\neq3$, where the constant $\tilde{\zeta}_{m}$ is
defined as
$\tilde{\zeta}_{m}=\zeta_{m}\left(\frac{2\tilde{K}_{m}}{3-m}\right)^{\frac{3(1-m)}{3-m}}$.

\begin{figure}[th]
{\hspace{-5
cm}\includegraphics[width=6.0in,angle=0,clip=true]{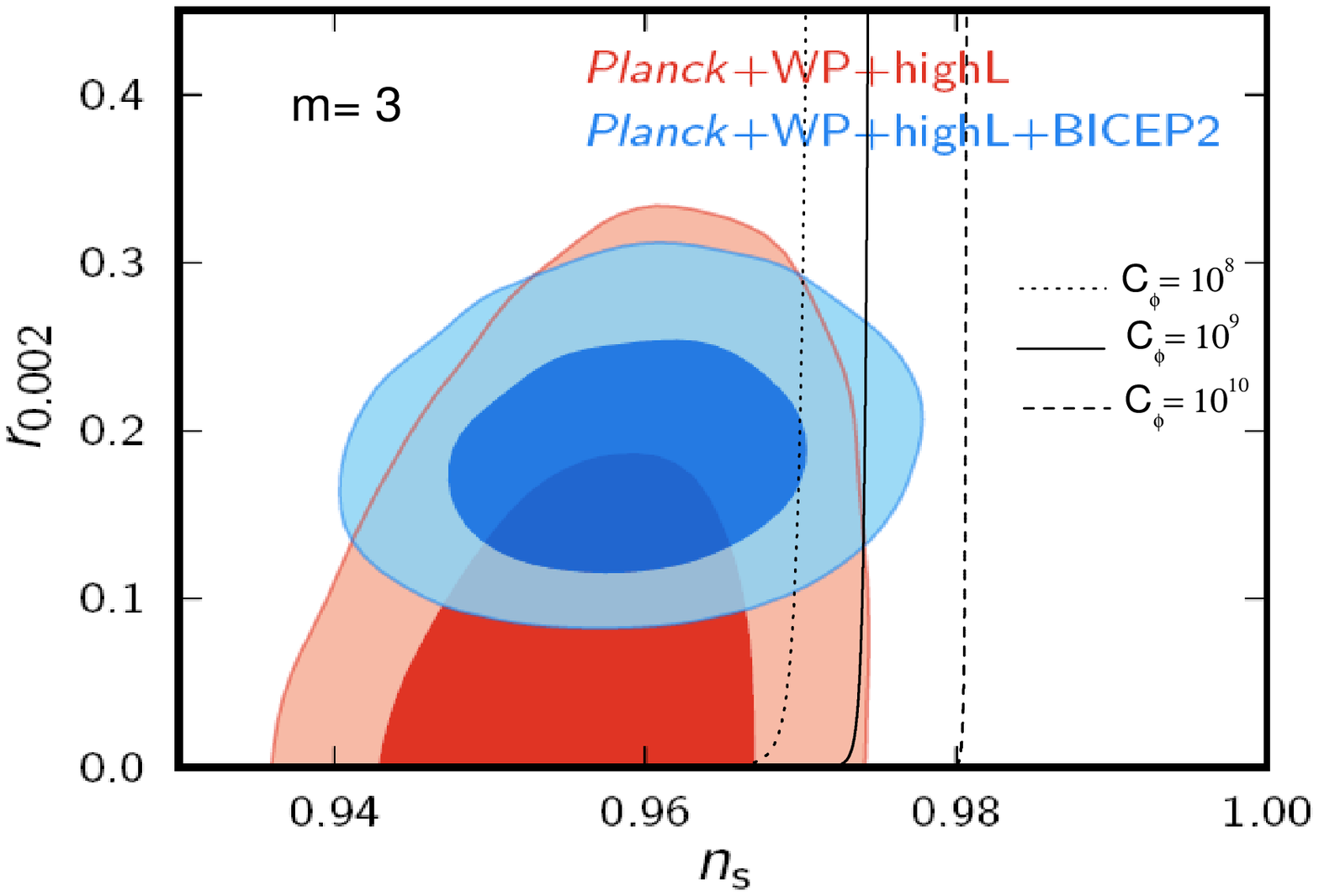}}
{\vspace{-3cm}\caption{Evolution of the tensor-scalar ratio $r$
versus the scalar spectrum index $n_s$ in the strong dissipative
regime, for three different values of the parameter $C_\phi$ and
$m=3$. The dotted, solid , and dashed lines are for the pairs
($A=4.86\times 10^{-5}$, $f=0.64$), ($A=3.92\times 10^{-6}$,
$f=0.82$) and ($A=1.08\times 10^{-6}$, $f=0.99$), respectively.
Also, in this plot we have taken the values $C_\gamma=70$,
$m_p=1$, $\mu=0.99$, $\eta=1$, $\epsilon=-1$, ${\cal{A}}_0=1$ and
$\Omega_{rc}=(\beta H_0^2)^{-1}=0.014$\cite{Azizi:2011ys}.
 \label{strong1}}}
\end{figure}

In Fig.(\ref{strong1}), we establish  the dependence of the
tensor-scalar ratio $r$ versus the primordial tilt $n_s$, for the
specific case in which we fix $m=3$ ($\Gamma\propto T^3/\phi^2$),
in the strong dissipative regime. Here, we have considered three
different values of the parameter $C_\phi$. Again, we show the
two-dimensional marginalized constraints (68$\%$ and 95$\%$ CL)
from BICEP2 experiment\cite{B2}. In order to write down values for
the the tensor-scalar ratio and the scalar spectrum index for the
special case $m=3$, i.e., $\Gamma\propto T^3/\phi^2$, we use Eqs.
(\ref{ns3N}) and (\ref{rfi3N}), where $C_\gamma=70$,  $m_p=1$,
$\mu=0.99$, $\eta=1$, $\epsilon=-1$, ${\cal{A}}_0=1$ and
$\Omega_{rc}=(\beta H_0^2)^{-1}=0.014$\cite{Azizi:2011ys}. Also,
we numerically make use of Eqs.(\ref{Prs3N}) and (\ref{ns3N}) and
obtain $A=1.08\times 10^{-6}$ and $f=0.99$ for the value of
$C_\phi=10^{8}$ for which $n_s=0.96$,
 ${\cal{P}_{\cal{R}}}=2.43\times10^{-9}$ and $N=60$. Similarly,
$C_\phi=10^{9}$ corresponds to $A=3.92\times 10^{-6}$, $f=0.82$;
for the case in which $C_\phi=10^{10}$, then $A=4.86\times
10^{-5}$, $f=0.64$. From the plot we observe  that the value
$C_\phi<10^{10}$ is well corroborated  by the CL from the BICEP2
experiment and also from Planck data. Additionally, we observe
that the parameter $C_\phi>10^{8}$ is well supported by the strong
regime, in which $\Gamma/3H> 1$ (not shown). In this form, the
range for the parameter $C_\phi$ in the specific case in which
$m=3$, is given by $ 10^{8}<C_\phi<10^{10}$. We note that this
range for the parameter $C_\phi$ in the strong dissipative regime,
becomes similar to the range obtained in Ref.\cite{9}. Finally,
for this case, we observed that we have found a lower bound for
the parameter $C_\phi$ from the ratio $\Gamma/3H>1$ and an upper
bound from BICEP2 experiment or Planck satellite.

For the case in which $m=1$ ($\Gamma\propto T$), we find that the
tensor-scalar ratio $r\simeq 0$, and the model is disproved from
BICEP2, since $r=0.2^{+0.07}_{-0.05}$, with $r=0$ disproved at
7.0$\sigma$.  However,  previous CMB observations  from the Planck
satellite and other CMB experiments
 yielded only an upper limit for the ratio $r<0.11$ (at 95$\%$ C.L.). In this form, for the case $m=1$  we numerically
 obtain  that the parameter $C_\phi>0.05$ is well supported by the strong regime, in which $\Gamma/3H >1$.
Also, we observe that when we increase the value of the parameter
$C_\phi$, the value of the tensor to scalar ratio $r\simeq\,0.$ In
particular, for the value $C_\phi=0.05$ corresponds to
$(\frac{\Gamma}{3H})\mid_{N=60}\simeq \,1.5$, the ratio
$(\frac{T}{H})\mid_{N=60}\simeq\,88 $  and the tensor to scalar
ratio $r\mid_{n_s=0.96}\simeq 0.002 $.

 Also, in the strong regime we observe that for the
cases in which $m=0$ and $m=-1$,  the models are disproved from
observations; since spectral index $n_s>1$, these
models do not work.

 Analogous to the case of the weak dissipative regime, we also
study the evolution of the ratio $T/H$ on the number of e-folds
$N$ for the strong dissipative regime. For the special case $m=3$,
we find that the constraint for the parameter $C_\phi$, given by $
10^{8}<C_\phi<10^{10}$ is well supported by the condition for the
warm inflation scenario in which $T/H>1$ when the number of
e-folds $N\geq 60$. Here, we numerically utilize Eqs.(\ref{at}),
(\ref{TF}), (\ref{phis}) and (\ref{Ns3}) together with the same
parameters of the Fig.(4). In particular,  for the value
$C_\phi=10^8$  we find that the value of ratio
$\left(\frac{T}{H}\right)\mid_{N=60}\simeq 49.5$, for the value
$C_\phi=10^9$ corresponds to
$\left(\frac{T}{H}\right)\mid_{N=60}\simeq 311.1$  and for
$C_\phi=10^{10}$ it corresponds to
$\left(\frac{T}{H}\right)|_{N=60}\simeq 1104.3$.

\section{Conclusions \label{conclu}}

In this paper we have studied the intermediate  inflationary model
in the context of warped DGP-warm inflation. In the slow-roll
approximation, we have obtained analytic solutions of the
equations of motion, during  the weak and strong regime, for a
general form of the dissipative coefficient. For the dissipative
coefficient $\Gamma(T,\phi)=C_\phi\,T^{m}/\phi^{m-1} $, we have
investigated the specific values $m=3$, $m=1$, $m=0$, and $m=-1$.
In our model, we have found  analytical  expressions for the
corresponding effective potential, power spectrum, scalar spectrum index,
and tensor to scalar ratio. From these quantities, we have
obtained, in both regimes, constraints on the parameters of the
model from the BICEP2 experiment and Planck, where we have consider the
constraint on the $r-n_s$ plane.

On the other hand,  we have obtained a constraint for the value of
the parameter $C_\phi$ from the both regimes, i.e.,  $\Gamma/3H<1$
or $\Gamma/3H>1$  and also we have found a new constraint for the
parameter $C_\phi$ from the  condition  the warm inflation
 $T>H$.  In the weak dissipative regime,
we have obtained an upper bound for the parameter $C_\phi$  from
BICEP2, Planck and a lower bound from the condition for the
warm inflation  $T>H$, and we have observed that when we decrease
the value of $m$, the value of the parameter $C_\phi$ also
decreases. For the strong dissipative regime, the model only works
for the case $m=3$, i.e., $\Gamma\propto T^{3}/\phi^2$. Here, we
have obtained a lower bound for the parameter $C_\phi$ from the
ratio $\Gamma/3H>1$, and an upper bound from the BICEP2
experiment and Planck data. For the case in which $m=1$ ($\Gamma\propto T$), we
have found that $r\simeq 0$, and the model is disproved by BICEP2.
  However, from the Planck satellite in which $r<0.11$, we have found a lower bound for the parameter $C_\phi$.
Finally, we have observed that for the cases in which $m=0$ and
$m=-1$, the models are disproved by observations, since the
spectral index $n_s>1$, and these models do not work.

\begin{acknowledgments}

The authors dedicate this article to the memory of Dr. Sergio
del Campo, mentor, colleague and friend. R.H. was supported by
COMISION NACIONAL DE CIENCIAS Y TECNOLOGIA through FONDECYT Grant
N$^0$ 1130628 and DI-PUCV N$^0$ 123.724. N. V. was supported by
Proyecto Beca-Doctoral CONICYT N$^0$ 21100261.

\end{acknowledgments}


\end{document}